\documentclass[prl,showpacs,twocolumn,amsmath,amssymb]{revtex4}

\usepackage{graphicx}
\usepackage{bm}

\begin{document}

\title{Long-range correlations and coherent structures in
magnetohydrodynamic equilibria}


\author{Peter B. Weichman}

\affiliation{BAE Systems, Advanced Information Technologies, 6 New
England Executive Park, Burlington, MA 01803}




\begin{abstract}

The equilibrium theory of the 2D magnetohydrodynamic equations is
derived, accounting for the full infinite hierarchies of conserved
integrals. An exact description in terms of two coupled elastic
membranes emerges, producing long-ranged correlations between the
magnetic and velocity fields. This is quite different from the results
of previous variational treatments, which relied on a local product
ansatz for the thermodynamic Gibbs distribution. The equilibria display
the same type of coherent structures, such as compact eddies and zonal
jets, previously found in pure fluid equilibria. Possible consequences
of this for recent simulations of the solar tachocline are discussed.

\end{abstract}

\pacs{47.10.-g, 
05.70.Ln,       
05.90.+m,       
52.30.-q}       

\maketitle

The ideal magnetohydrodynamic (MHD) equations
\begin{eqnarray}
\partial_t {\bf v} + ({\bf v} \cdot \nabla) {\bf v}
&=& -\nabla p + {\bf J} \times {\bf B}
\nonumber \\
\partial_t {\bf B} &=& \nabla \times ({\bf v} \times {\bf B})
\label{1}
\end{eqnarray}
describe the evolution the velocity field ${\bf v}$ of a perfectly
conducting fluid under the influence of pressure gradients and the
magnetic Lorentz force, and advection of the magnetic field ${\bf B}$
by the velocity field \cite{MHDref}. The equations are closed through
Ampere's law ${\bf J} = \nabla \times {\bf B}$ and the
incompressibility constraints $\nabla \cdot {\bf v} = 0$, $\nabla \cdot
{\bf B} = 0$.

A number of systems such as the solar tachocline \cite{TDH2007}, and
possibly the Earth's core-mantle boundary \cite{B2007}, are
approximately governed by a 2D approximation in which ${\bf v},{\bf B}$
are horizontal, depending only on the horizontal coordinates ${\bf r} =
(x,y)$ in a domain $D$ of the $xy$-plane (with 1D boundary $\partial
D$) \cite{TDM2011}. The current ${\bf J} = J {\bf \hat z}$ and
vorticity $\nabla \times {\bf v} = \omega {\bf \hat z}$ may then be
treated as scalars. One may also express ${\bf B},{\bf v}$ in terms of
a potential $A$ and stream function $\psi$: ${\bf B} = \nabla \times (A
{\bf \hat z})$, ${\bf v} = \nabla \times (\psi {\bf \hat z})$, with the
2D Laplacian relationships $J = -\nabla^2 A$, $\omega = -\nabla^2
\psi$. The equations then reduce to two scalar equations
\begin{eqnarray}
\partial_t \omega + {\bf v} \cdot \nabla (\omega + f)
&=& {\bf B} \cdot \nabla J
\nonumber \\
\partial_t A + {\bf v} \cdot \nabla A &=& 0
\label{2}
\end{eqnarray}
in which the Coriolis parameter $f({\bf r}) = 2 \Omega \sin(\varphi)$,
with latitude $\varphi(y)$, accounts frame of reference rotating with
angular velocity $\Omega$. In the beta-plane approximation one
linearizes $f(y) = f_0 + \beta y$ about some reference $\varphi_0$. The
total energy is given by
\begin{equation}
E = \frac{1}{2} \int d^2r \left(|{\bf v}|^2 + |{\bf B}|^2 \right)
= \frac{1}{2} \int d^2r \left(|\nabla \psi|^2 + |\nabla A|^2 \right).
\label{3}
\end{equation}
Its conservation requires lossless boundary conditions, e.g., periodic
or ``free-slip'': ${\bf B} \cdot {\bf \hat n} = 0$, ${\bf v} \cdot {\bf
\hat n} = 0$, with ${\bf \hat n}$ the local normal to $\partial D$. The
latter imply constant values of $A,\psi$ on each connected component of
$\partial D$.

The pure 2D fluid equation (${\bf B} \equiv 0$) exhibits a turbulent
inverse cascade: small scale vortices self-organize into large scale
flows, limited only by the domain size. In a forced system, this leads
to growing flows, limited only by dissipation. These take the form of
strong jet-like structures, or large coherent vortices, which have
strong implications for geophysical flow stability and global
transport. The weather bands and Great Red Spot of Jupiter are famous
examples. For weak driving and dissipation, these flows may be modeled
as near-equilibrium. Application of statistical mechanics to the fluid
equations indeed produces such structures
\cite{M1990,RS1991,MWC1992,MR1994,WP2001,BS2002,W2006,BV2012}.

Similar flows in the solar tachocline, which sharply divides the
rigidly rotating interior radiation zone from the differentially
rotating outer convection zone, would have strong implications for
angular momentum transport between the two zones \cite{TDH2007}. Recent
2D MHD simulations \cite{TDH2007,TDM2011}, however, have found that the
presence of even weak ${\bf B}$ tends to break up large scale flows.
Although ${\bf B}$ indeed destroys the infinite set of vorticity
conservation laws that lead to the fluid inverse cascade, they are
replaced by an entirely new infinite set of magnetic conservation laws
\cite{HMRW1985}. This motivates the present investigation of the
equilibria allowed by these new laws, and what conditions might limit
their amplitude.

It will be shown that the new conservation laws indeed produce large
scale flows, but of distinctly different character. In contrast to
Euler flow, where all initial energy cascades to large scales, in 2D
MHD a finite fraction flows to small scales, generating large
microscale fluctuations. For small initial ${\bf B}$, these can be much
larger than the mean flow \cite{TDH2007,TDM2011}. However, dissipation
effects differentially tend to suppress these, and some form of driving
might amplify the mean flow to more visible levels. The results here
may then point to more careful studies of the types driving and initial
conditions that could produce stronger large scale flows, and their
physical realizability. In addition, unlike Euler flow, in which the
small scale fluctuations are completely uncorrelated on the microscale
\cite{M1990,MWC1992}, the 2D MHD flows are shown to exhibit long-ranged
power-law correlations, that could also be confirmed in simulations.
Such correlations were invisible to earlier treatments which used a
local product approximation for the Gibbs distribution
\cite{JT1997,LDC2005}.

The MHD equations are nonlinear, thereby producing turbulent initial
evolution from a spatially complex initial condition. However, at late
time, the flow may equilibrate to a steady state. Unusually, such 2D
fluid steady states are not necessarily quiescent, but can exhibit
spatial structure such as large scale vortices or zonal jets
\cite{M1990,RS1991,MWC1992,MR1994,WP2001,BS2002,W2006,BV2012}. The
origin, and diversity, of these structures lies in the infinite number
of conservation laws, beyond the usual energy (along with momentum or
angular momentum if the relevant translation or rotation symmetry
exists) constraining the flow \cite{HMRW1985}. If one sets ${\bf B}
\equiv 0$, the first line of (\ref{2}) is the Euler equation, and the
resulting advective conservation of the potential vorticity $\omega_p =
\omega + f$ implies conservation of all spatial integrals (Casimirs)
$K_g = \int_D d^2 g(\omega_p) d^2 r$, with $g$ an arbitrary 1D
function. However, any nonzero ${\bf B} \cdot \nabla J$ breaks this
conservation \cite{foot:eulereqm}, replacing it by advective
conservation of $A$ [second line of (\ref{2})]. The new Casimirs are
\cite{HMRW1985}
\begin{equation}
J_g = \int d^2r g(A),\
K_g = \int_D d^2r (\omega + f) g(A)
\label{4}
\end{equation}
for arbitrary 1D functions $g$. Integrating by parts, one may also
replace $\omega g(A) \to ({\bf v} \cdot {\bf B}) g'(A)$. These may be
parameterized as conservation of the functions $j(\sigma),k(\sigma)$
for all $\sigma$ obtained using $g(s) = \delta(s-\sigma)$. Finally, any
conserved momentum may be expressed in the form
\begin{equation}
P = \int d^2r \alpha \omega
= \int d^2 r (\nabla \alpha) \times {\bf v}
= -\int d^2 r \nabla \alpha \cdot \nabla \psi,
\label{5}
\end{equation}
for some fixed function $\alpha({\bf r})$ \cite{foot:momentum}.

Under the usual ergodic assumptions (whose validity is far from
obvious, and known to be violated for some initial conditions
\cite{CC}), the equilibrium statistics are obtained from the grand
canonical partition function
\begin{equation}
Z[\beta,\lambda,\mu,\nu]
= \int D[A,\psi] e^{-\beta {\cal G}[A,\psi]},
\label{6}
\end{equation}
and associated free energy ${\cal F} = -T \ln(Z)$. The functional
integral is over all fields $A,\psi$ (obeying the appropriate boundary
conditions), $\beta = 1/T$ is the inverse temperature, and the Gibbs
functional is ${\cal G} = E - J_\mu - K_\nu - \lambda P$:
\begin{eqnarray}
{\cal G} &=& \int d^2r \bigg\{\frac{1}{2}|\nabla A|^2
+ \frac{1}{2}|\nabla \psi|^2 - \nu'(A) \nabla A \cdot \nabla \psi
\nonumber \\
&&\hskip0.5in +\ \lambda \nabla \alpha \cdot \nabla \psi
- [\mu(A) + f \nu(A)]\bigg\},
\label{7}
\end{eqnarray}
where Lagrange multipliers have been introduced for each conserved
integral: functions $\mu(s)$, $\nu(s)$ and parameters $\beta$,
$\lambda$ are adjusted to obtain the values of $j(\sigma)$,
$k(\sigma)$, $P$, $E$ defined by the initial flow. The physical model
associated with ${\cal G}$ is that of two membranes, with ``heights''
$A,\psi$, and unit surface tension, coupled through their gradients.
The $P$ term also acts to bias $\nabla \psi$; in particular, ${\cal G}$
favors ${\bf B}$ parallel to $\nu'(A) {\bf v} + \lambda \nabla \times
\alpha$. The term $\mu(A)+f\nu(A)$ is an external potential, confining
$A$ near its minimum, and depends (smoothly) on position through
$f({\bf r})$.

The equilibria governed by ${\cal G}$ have been previously investigated
\cite{JT1997,LDC2005} using a variational approach in which ${\bf
v},{\bf B}$ were treated as spatially uncorrelated. The exact physics
of ${\cal G}$, however, dictates something quite different. The elastic
interactions generate very long-ranged correlations, with distortion in
the surfaces interacting in a Coulomb-like fashion, leading to
log-divergent fluctuations, and dipole-like correlations for ${\bf
v},{\bf B}$.

To understand the nature of the equilibrium states, it is extremely
useful begin with a discrete spatial mesh, and then consider the
continuum limit. To this end, using a square lattice with mesh size $a$
one obtains
\begin{eqnarray}
\beta {\cal G}_a &=& \frac{1}{2} \beta \sum_{\langle i,j \rangle}
\big\{(A_i - A_j)^2 + (\psi_i - \psi_j))^2
\nonumber \\
&&-\ [\nu'(A_i) + \nu'(A_j)] (A_i - A_j) (\psi_i - \psi_j)
\nonumber \\
&&+\ 2 \lambda (\alpha_i - \alpha_j)
(\psi_i - \psi_j) \big\}
\nonumber \\
&&-\ \beta a^2 \sum_i [\mu(A_i) + f_i \nu(A_i)],
\label{8}
\end{eqnarray}
in which $\langle i,j \rangle$ are nearest neighbors. The functional
integral $D[A,\psi] \to \prod_i \int dA_i d\psi_i$ now becomes (up to
overall normalization) an independent product \cite{FM1976}. We define
also ${\bf B}_i = a^{-1} (A_{i+{\bf \hat y}} - A_i, A_i-A_{i+{\bf \hat
x}})$, ${\bf v}_i = a^{-1} (\psi_{i+{\bf \hat y}} - \psi_i,
\psi_i-\psi_{i+{\bf \hat x}})$. It is apparent here that if $\beta$
remains finite as $a \to 0$ then the first term will yield
$O(\sqrt{T})$ fluctuations between neighboring sites, and the second
term becomes negligible \cite{foot:infmunu}. Well defined hydrodynamic
equilibria, with nontrivial competition between kinetic and potential
energies requires that $\beta$ scale with $a$. Specifically, taking
$\beta(a) = \bar \beta/a^2$ yields $O(a {\bar T}^{1/2})$ site-to-site
fluctuations, and continuous $A({\bf r}), \psi({\bf r})$. On the other
hand ${\bf B}_i$, ${\bf v}_i$ will have finite, $O(\sqrt{\bar T})$,
site-to-site fluctuations, yielding non-differentiable continuum
$A,\psi$. With this scaling there are two surviving contributions to
${\cal G}$ as $a \to 0$. Decomposing $A = A_0 + \delta A$, $\psi =
\psi_0 + \delta \psi$ in which $A_0 = \langle A \rangle$, $\psi_0 =
\langle \psi \rangle$ are the equilibrium averages (self-consistently
determined below) one obtains
\begin{eqnarray}
{\cal G}[A,\psi] &=& {\cal G}[A_0,\psi_0]
+ {\cal G}_2[\delta A,\delta \psi;A_0] + O(a),
\nonumber \\
{\cal G}_2 &=& \frac{1}{2} \int d^2r \big[|\nabla \delta A|^2
+ |\nabla \delta \psi|^2
\nonumber \\
&&\hskip0.3in -\ 2\nu'(A_0) (\nabla \delta A)
\cdot (\nabla \delta \psi) \big]
\nonumber \\
&=& \frac{1}{2} \int d^2r
\left[[1+\nu'(A_0)]|\nabla \delta \phi^-|^2 \right.
\nonumber \\
&&\hskip0.3in \left.+\ [1-\nu'(A_0)]
|\nabla \delta \phi^+|^2 \right],
\label{9}
\end{eqnarray}
in which $\delta \phi^\pm = (\delta A \pm \delta \psi)/\sqrt{2}$ are
independent Gaussian fields \cite{foot:nuprime}. We have reverted, for
compactness, to continuum notation. For smooth $A_0,\psi_0$, all other
terms, including those linear in $\delta A, \delta \psi$, vanish with
$a \to 0$. From ${\cal G}_2$ one therefore obtains the exact free
energy
\begin{equation}
{\cal F}[\bar\beta,\lambda,\mu,\nu]
= {\cal G}[A_0,\psi_0] + {\cal F}_2[A_0]
\label{10}
\end{equation}
in which the Gaussian fluctuation free energy ${\cal F}_2$ has a well
defined $a \to 0$ limit. It is not computable in closed form for
general $A_0$, but for constant $\nu'(A_0) \equiv \nu_0'$ and periodic
boundary conditions one obtains
\begin{eqnarray}
{\cal F}_2 &=& \frac{|D|}{\bar \beta} \int_\mathrm{BZ}
\frac{d^2k}{(2\pi)^2} \ln\left[\frac{\bar \beta}{2\pi}
\sqrt{1-\nu_0^{\prime 2}} E({\bf k}) \right]
\nonumber \\
E({\bf k}) &\equiv& 4 \left[\sin^2(k_x/2) + \sin^2(k_y/2) \right],
\label{11}
\end{eqnarray}
in which $|D|$ is the area of $D$, and the Brillouin zone is defined by
$\pi < k_x,k_y \leq \pi$. More generally, ${\cal F}_2$ is obtained from
the log-sum of the eigenvalues of the (generalized Laplacian) operators
${\cal O}^\pm = -\nabla \cdot [1 \mp \nu'(A_0)] \nabla$.

The divergence of $\beta = \bar \beta/a^2$ in (\ref{6}) means that
$A_0,\psi_0$ are determined self-consistently by minimizing ${\cal F}$.
The conditions $\delta {\cal G}_0/\delta \psi_0({\bf r}) = 0$, $\delta
({\cal G}_0 + {\cal F}_2)/\delta A_0({\bf r}) = 0$ produce,
respectively, the equilibrium equations
\begin{eqnarray}
{\bf v}_0 &=& \nu'(A_0) {\bf B}_0 - \lambda \nabla \times \alpha
\nonumber \\
\omega_0 + \nu'(A_0) J_0 &=& \mu'(A_0) + f \nu'(A_0)
\nonumber \\
&&+\ \nu''(A_0) \gamma({\bf r},{\bf r};A_0),
\label{12}
\end{eqnarray}
where ${\bf B}_0 = \nabla \times (A_0 {\bf \hat z})$, etc., are the
other derived equilibrium fields. The first equation provides a direct
relation between the equilibrium velocity and magnetic field: ${\bf
B}_0$ is colinear with ${\bf v}_0$ up to a mean flow subtraction.
Substituting the curl of the first equation into the second, one
obtains a closed equation for $A_0$:
\begin{eqnarray}
&&-\nabla \times \{[1- \nu'(A_0)^2]{\bf B}_0 \}
= \mu'(A_0) + f \nu'(A_0) + \lambda \nabla^2 \alpha
\nonumber \\
&&\hskip0.6in +\ \frac{1}{2}\left[\nu'(A_0)^2 \right]' |{\bf B}_0|^2
+ \nu''(A_0) \gamma({\bf r},{\bf r}).
\label{13}
\end{eqnarray}
The fluctuation-derived nonlocal term at the end is obtained from the
cross-correlation function
\begin{eqnarray}
\gamma({\bf r},{\bf r}') = \langle \nabla \delta A({\bf r})
\cdot \nabla \delta \psi({\bf r}') \rangle_2
= \langle \delta {\bf B}({\bf r})
\cdot \delta {\bf v}({\bf r}') \rangle_2.
\label{14}
\end{eqnarray}
If one defines the Green functions of the operators ${\cal O}^\pm$,
\begin{equation}
-\nabla \cdot [1 \mp \nu'(A_0)] \nabla G^\pm({\bf r},{\bf r}')
= \delta({\bf r}-{\bf r}'),
\label{15}
\end{equation}
which characterize the membrane fluctuations,
\begin{equation}
G^\pm({\bf r},{\bf r}') = \frac{\bar\beta}{a^2}
\langle [\delta \phi^\pm({\bf r})
-\delta \phi^\pm({\bf r}')]^2 \rangle_2,
\label{16}
\end{equation}
then one obtains the relation
\begin{equation}
\gamma({\bf r},{\bf r}') = \frac{1}{2} \nabla \cdot \nabla'
[G^+({\bf r},{\bf r}') - G^-({\bf r},{\bf r}')].
\label{17}
\end{equation}
Solutions to (\ref{15}) yield the electrostatic potential generated by
a unit charge at ${\bf r}'$ with spatially varying dielectric function
$\epsilon^\pm({\bf r}) = 1 \mp \nu'(A_0)$. By Gauss's law, the
displacement field $\epsilon^\pm \nabla G^\pm$ has unit flux through
any bounding contour, and it follows that $G^\pm({\bf r}-{\bf r}') \sim
(1/2\pi \epsilon^\pm) \ln(|{\bf r}-{\bf r}'|/a)$ grows logarithmically
with distance. This unbounded wandering reflects the usual thermal
roughening result for 2D membranes \cite{foot:wander}.

\begin{figure}
\includegraphics[width=1.65in,bb=118 303 480 495,clip]{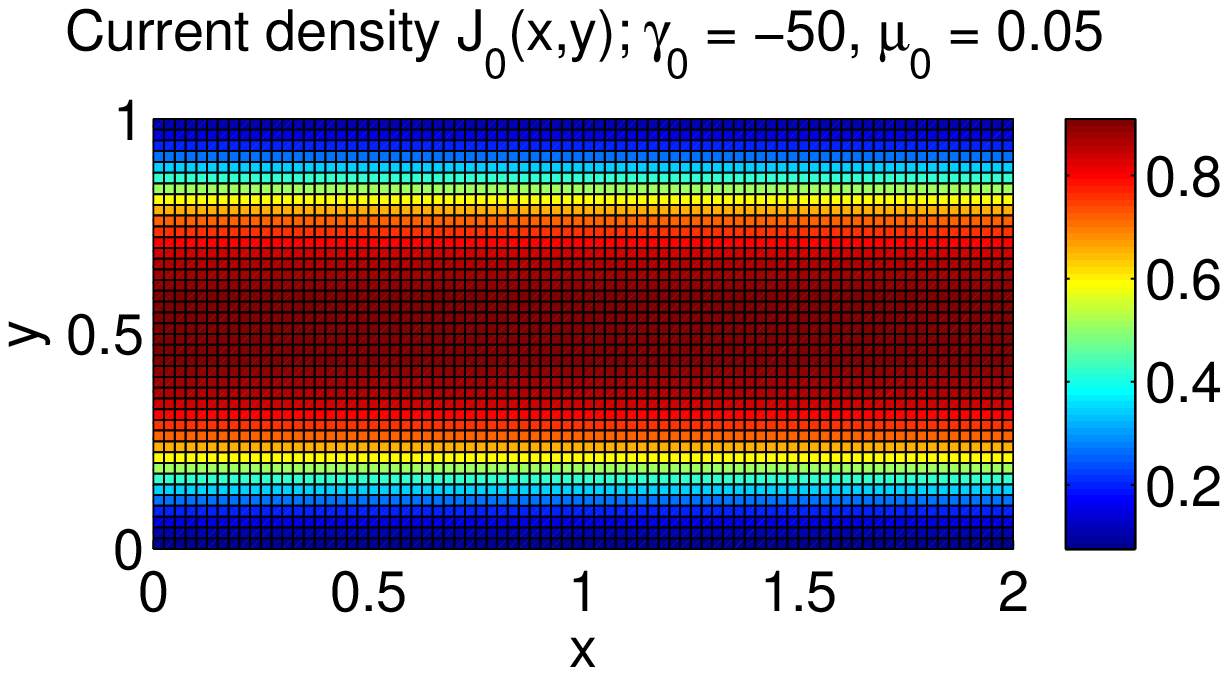}
\includegraphics[width=1.65in,bb=118 303 480 495,clip]{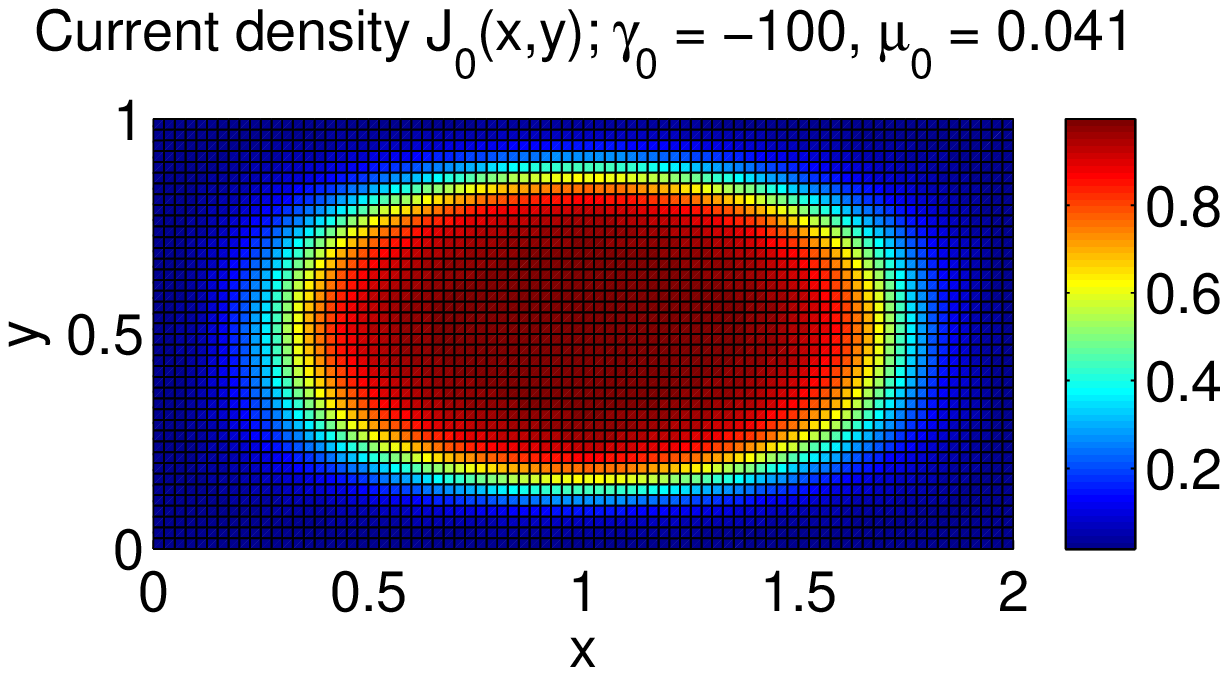}

\includegraphics[width=1.65in,bb=118 303 480 495,clip]{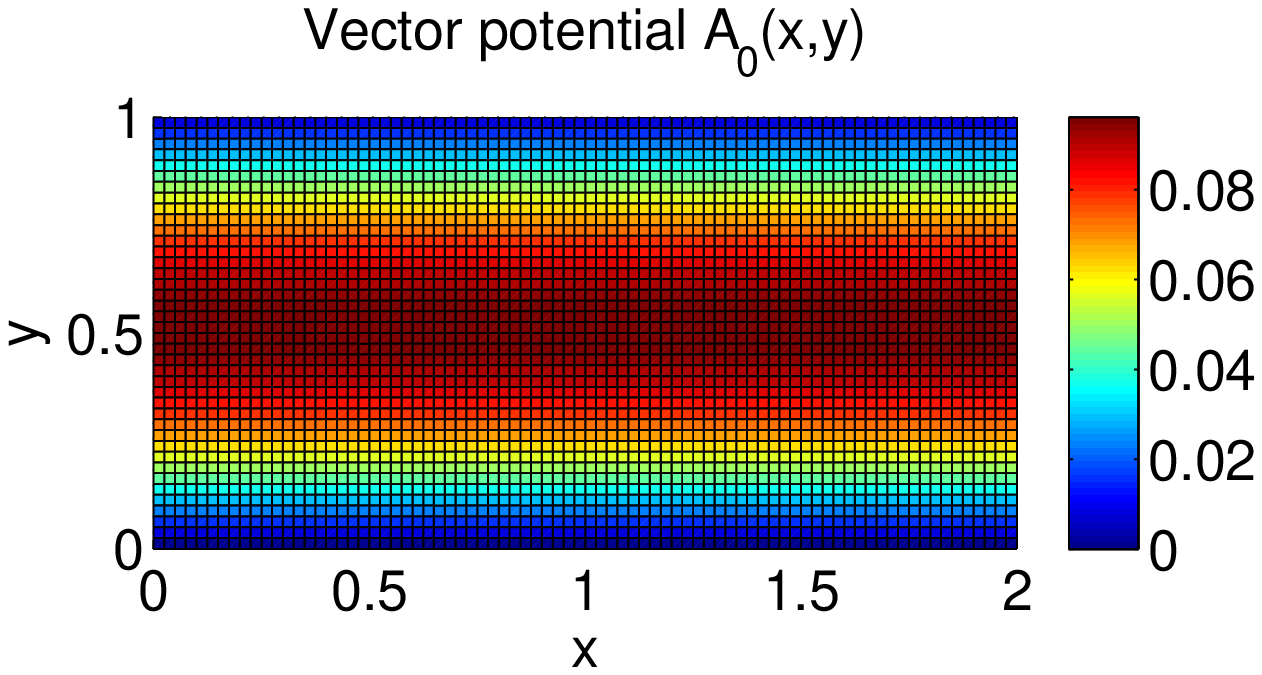}
\includegraphics[width=1.65in,bb=118 303 480 495,clip]{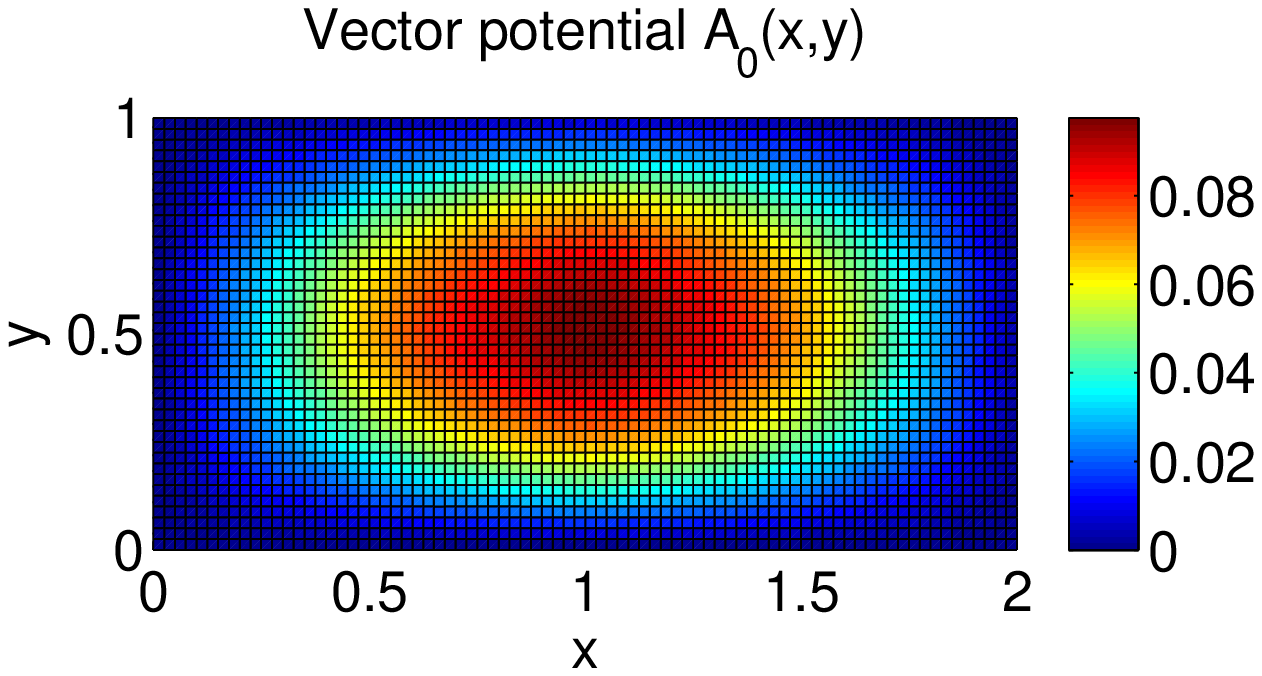}

\caption{\textbf{Left:} Equilibrium jet solution obtained using
periodic boundary conditions in $x$. \textbf{Right:} Equilibrium vortex
solution obtained using free slip boundary conditions in both $x$ and
$y$. Vortex solutions exist as well for periodic boundary conditions,
but the numerics are more involved. Stream lines and magnetic field
lines are level curves of constant potential $A_0$ (bottom panels) or
current density $J_0$ (top panels). For simplicity we set $\bar T = 0$
(no microscale fluctuations), $f = 0$ (no rotation), $\lambda = 0$ (no
net mean flow), take constant $\nu'(A) = \nu_0'$ (hence ${\bf B}_0 =
{\bf v}_0/\nu_0'$), and use the form $\mu(A) = -\gamma^{-1} \ln[1 +
e^{-\gamma(A-\mu_1)}]$. For large $\gamma < 0$, the latter tends to
produce a two level system, with $J_0 = \omega_0/\nu_0'$ taking values
0 or $q = (1-\nu_0^{\prime 2})^{-1}$, and the parameter $\mu_1$
controls the proportion of each level. Plotted are actually the scaled
quantities $A_0/q$, $J_0/q$, with parameters $\gamma_0 = \gamma q$,
$\mu_0 = \mu_1 q$ specified in the titles. The results are then
independent of $q$ (and hence of the choice of $\nu_0'$).}

\label{fig:eqmsolns}
\end{figure}

For a translation invariant system (i.e., periodic, uniform), one may
replace $\nabla' \to -\nabla$ and (\ref{15}) implies that $\gamma
\equiv 0$ for ${\bf r} \neq {\bf r}'$, and $\gamma({\bf r},{\bf r}) =
\nu_0'/2\bar \beta [1- \nu_0'^2]$. For this case, the locality
assumption \cite{JT1997} is valid \cite{foot:vdotb}. Otherwise one
obtains nonlocal contributions to $\gamma({\bf r},{\bf r})$, even for a
uniform system with free slip boundary conditions: nonlocal
contributions then arise from image charges outside $D$ that enforce
the boundary conditions on $G^\pm$.

Once the equilibrium solution is obtained, the values of the conserved
variables may be obtained from derivatives of ${\cal F}$ at fixed
$A_0,\Psi_0$:
\begin{eqnarray}
E &=& \frac{1}{2} \int_D d^2r
\left[|{\bf v}_0|^2 + |{\bf B}_0|^2
+ \varepsilon({\bf r},{\bf r}) \right]
\nonumber \\
P &=& -\frac{\partial \cal F}{\partial \lambda}
= \int_D d^2r \alpha \omega_0
\nonumber \\
j(\sigma) &=& -\frac{\delta \cal F}{\delta \mu(\sigma)}
= \int_D d^2r \delta[\sigma - A_0({\bf r})]
\nonumber \\
k(\sigma) &=& -\frac{\delta \cal F}{\delta \nu(\sigma)}
= \int_D d^2r \big\{(\omega_0 + f)\delta[\sigma - A_0({\bf r})]
\nonumber \\
&&\hskip0.75in -\ \gamma({\bf r},{\bf r})
\delta'[\sigma - A_0({\bf r})] \big\}.
\label{18}
\end{eqnarray}
where $\varepsilon({\bf r},{\bf r}') = -\nabla \cdot \nabla'[G^+({\bf
r},{\bf r}') + G^-({\bf r},{\bf r}')]$ determines the microscale
fluctuation part of the energy. The extremum condition ensures that all
derivatives with respect to the implicit dependence of $A_0,\psi_0$ on
the Lagrange multipliers cancels out.

Comparing (\ref{4}), it is seen, due to continuity of $A$, that
$j(\sigma)$ is a large-scale quantity: it is its own equilibrium
average. Hence the level sets of the initial $A({\bf r},t=0))$ are
exactly preserved in the equilibrium $A_0({\bf r})$. One implication is
that a low-amplitude initial $A$ will produce an identically
low-amplitude $A_0$. However, e.g., for a spatially irregular initial
condition with comparatively large energy, one will have $\varepsilon
\gg |{\bf v}_0|^2,{\bf B}_0|^2$ and the physical equilibrium fields
will be masked by fluctuations. This may help explain what is observed
in simulations \cite{TDH2007,TDM2011}. For increasing initial $A$,
especially in the presence of forcing and weak dissipation, one may
expect clear equilibria to emerge more strongly, but the exact
conditions required for this remain to be determined.

Equations (\ref{12}), (\ref{17}) and (\ref{18}) are the basic results
of this paper. These equations are nonlocal and highly nonlinear, and
generally require a numerical solution. However, some general
properties may be inferred from the basic form of the effective
Hamiltonian (\ref{9}). The ${\cal G}_0$ term reflects a straightforward
classical surface tension minimization problem. The fluctuations
locally stretch the membranes and hence act to renormalize the surface
tension, but in a way the depends self-consistently on the average
membrane position $A_0,\psi_0$, which in turn respond to the external
forces provided by the $\mu + f\nu$ and $\lambda \alpha$ terms in
(\ref{7}). Thus, a stretched region of a membrane may be expected to
have reduced amplitude fluctuations. The minimization of ${\cal F}$
accounts fully for both effects. Figure \ref{fig:eqmsolns} shows
examples of very simple equilibrium solutions in a rectangular domain
(with model parameters specified in the caption), demonstrating the
existence of both vortex and jet structures.

Unlike the Euler case, where the conserved integrals ensure bounded
$\omega = \nabla \times {\bf v}$, thereby generating continuous ${\bf
v}$ and differentiable $\psi$, for 2D MHD only continuity of $A,\psi$
are provided, and this occurs now in response to the squared-gradients
in the energy, not in response to the conserved integrals. The latter
also ensures bounded $\delta {\bf v},\delta {\bf B}$, but this in turn
allows for finite microscale energy density $\varepsilon$. The second
derivatives $\omega,J$ then have unbounded fluctuations. Since
simulations often propagate $A,\omega$ using (\ref{2}), extra care may
then be required to ensure reasonable equilibration. In particular, if
(e.g., hyperviscous) dissipation acts too strongly to quell the
micro-fluctuations, it may also bleed energy out of the large scale
flow. There could also be physical analogues of this effect, depending
on the precise nature, e.g., of the true solar tachocline dissipation
mechanisms. These, and presumably many other, considerations must enter
the implications of the theory developed here.

\emph{Acknowledgments:} This material is based upon work supported in
part by the National Science Foundation Grant No.\ 1066293 and the
hospitality of the Aspen Center for Physics. The author has greatly
benefited from discussions with A. M. Balk, J. B. Marston, and J. Cho.


\begin{thebibliography}{}
\bibitem{foot:infmunu} More precisely, for bounded initial condition,
    $\mu(\sigma)$ or $\nu(\sigma)$ will be infinite outside a
    corresponding bounded set of $\sigma$. This will confine the
    membranes between infinite barriers, but their fluctuations will be
    otherwise free.
\end{thebibliography}
\end{document}